\documentclass[a4paper]{article}

\usepackage{icrc2013}

\title{The Formation of Condensation on Cherenkov Telescope Mirrors}

\shorttitle{Condensation on IACT Mirrors}

\authors{
P.M.Chadwick$^{1}$,
S.A. Cleaver$^{1,5}$,
M.Dyrda$^{2}$,
A. F\"orster$^{3}$,
J. Michalowski$^{2}$,
J. Niemiec$^{2}$,
C. Schultz$^{4}$,
M. Stodulski$^{2}$
for the CTA Consortium.
}

\afiliations{
$^1$ University of Durham, UK \\
$^2$ Institute of Nuclear Physics PAS, Krakow, Poland \\
$^3$ Max Planck Institut f\"ur Kernphysik, Heidelberg, Germany \\
$^4$ Universita di Padova and INFN, Italy \\
\scriptsize{
$^{5}$ now at: Astrium Ltd. \\
}
}

\email{p.m.chadwick@durham.ac.uk}

\abstract{The mirrors of imaging atmospheric Cherenkov telescopes are different from those of conventional astronomical telescopes in several ways, not least in that they are exposed to the elements. One of the issues which may arise is condensation forming on the mirrors during observing under certain atmospheric conditions, which has important consequences for the operation of the telescopes. This contribution discusses why telescope mirrors suffer condensation and describes the atmospheric conditions and mirror designs which are likely to be problematic.}


\begin{document}
\maketitle

\section{Introduction}

Imaging atmospheric Cherenkov telescopes (IACTs) require mirrors that are robust, lightweight and have high reflectivity in the blue/UV region of the spectrum. Yet these mirrors must also be inexpensive and capable of rapid manufacture due to the large mirror areas required. This unusual combination of properties has led to many innovative approaches to mirror production, including the use of composite structures and materials such as fibreglass in addition to the more traditional solid glass mirrors (see, e.g. \cite{bib:forster1} for a summary). Recently, the requirement for a robust surface that is highly reflective in the blue region of the electromagnetic spectrum and also has low reflectivity in the red (in order to reduce the effects of sky background) has led to research into dielectric-coated materials, with layers of e.g. SiO$_2$ and Ta$_2$O$_5$ \cite{bib:forster2}. These can be tuned to provide the required performance, and have the further advantage of being particularly resistant to abrasion.\\

No mirror type is without problems; in particular, although providing excellent optical quality, solid glass mirrors are heavy and time-consuming to produce. Composite mirrors, usually consisting of an aluminium honeycomb substrate bonded to a pre-formed reflective surface, are particularly useful in this respect, being generallly $\sim 50$\% lighter than solid glass mirrors for a given area. However, anecdotal evidence from current and previous IACT experiments is that composite mirrors are particularly prone to condensation on their surface during observations (see, e.g. \cite{bib:brazier}). This causes absorption and scattering of the Cherenkov light and results in a considerable drop in the telescopes' count rate. We report here the results of a study of this phenomenon, undertaken in the light of the design of mirrors for the forthcoming Cherenkov Telescope Array (CTA).\\

\section{The Formation of Condensation}

The likelihood of condensation forming on a mirror surface depends on the ambient temperature, the temperature of the surface and the relative humidity. For detailed studies of the formation of condensation, see \cite{bib:barry, bib:marsh}; the following is a basic account.

The temperature at which air becomes saturated, and will therefore form condensation, is the dew point, $T_d$. This can be determined from the values of the ambient temperature, $T_{amb}$ and the relative humidity $RH$ by employing a variant of the Magnus-Tetens formula \cite{bib:barenburg}:

\begin{equation}
T_d = \frac{b\phi}{a-\phi}
\end{equation}
where:

\begin{equation}
\phi = \frac{aT_{amb}}{b + T_{amb}} +ln\left(\frac{RH}{100}\right)
\end{equation}

with the Magnus coefficients defined as $a$ = 17.27 and $b = 237.7~^{\circ}C$. As soon as the mirror surface temperature is equal to the dew point temperature, moist, warmer air in contact with the mirror surface will cool to the dew point and form condensation.

In sub-zero temperatures it is possible for frost to form on a surface without condensation forming first. In this case, the stronger bonding in solid water raises the frost point \textit{above} the dew point. Fitting to data from dew point/frost point conversion tables \cite{bib:bry}, we find that the frost point $T_f$ is related to $T_d$ as follows:

\begin{equation}
T_f = \alpha T_d + \beta
\end{equation}

with $\alpha = 0.90 \pm 0.01$ and $\beta = 0.1 \pm 0.2$. This equation is only valid when $T_d < 0~^{\circ}C$.

\section{The Cooling of Surfaces}

In order to model the cooling on any surface, conduction, convection and radiation must be taken into account.

\subsection{Conduction}

It is likely that the reflective, front surface of a mirror will possess a lower surface temperature than the rear of the mirror, due to radiative cooling of the reflective surface when it is pointed at the night sky, and therefore a temperature gradient will exist from the rear to the front of the mirror. The rate of thermal energy transfer in Watts can be expressed as:

\begin{equation}
\frac{dQ}{dt} = \frac{kA(\Delta T)}{L} \label{eq:cond}
\end{equation}

where $k$ is the thermal conductivity of the mirror in $\rm{Wm}^{-1}\rm{K}^{-1}$ and $L$ its thickness, $A$ is the area of the material in m$^2$ and $\Delta T$ is the temperature difference between the front and the rear of the mirror.

\subsection{Convection}

Convective heat transfer is likely via forced rather than natural convection, as winds will flow over the mirror surface. The rate of heat transfer bewteen the mirror surface and the ambient air is given by:

\begin{equation}
\frac{dQ}{dt} = -hA(\Delta T)
\end{equation}

where $h$ is the convective heat transfer coefficient in $\rm{Wm}^{-1}\rm{K}^{-1}$ and $\Delta T$ in this case is the difference between the ambient air temperature and that of the mirror surface. In this study, $h$ was taken to be $(5.7 + 3.8v)$ where $v$ is the wind speed passing across a flat surface in $\mathrm{m}\;\mathrm{s}^{-1}$ \cite{bib:duffie}.

\subsection{Radiation}

The most important thermal process for the purpose of this study is radiative heat transfer. During observations, IACT mirrors are pointed towards the clear night sky, which is partially transparent in the 8 to 14 $\mu$m waveband \cite{bib:berdahl}. The sky therefore has an effective temperature considerably lower than the ambient temperature, and the mirrors will, as far as possible, equalise with this low temperature by the loss of thermal radiation, particularly in the 8 to 14 ~$\mu$m waveband. It has been shown that a typical black body surface will drop to temperatures of 18 to 33~$^{\circ}C$ below the ambient temperature by radiative cooling alone \cite{bib:eriksson}. However, any interfering medium, particularly a high concentration of water vapour, will inhibit the transfer of radiation. Radiation reflects off the interfering medium and produces a counteracting `down-welling' radiation. This is related to the emissivity of the night sky between 8 and 14 $\mu$m, $\epsilon_{sky}$, which in turn is directly related to the relative humidity of the ambient air and can be expressed as \cite{bib:idso}:

\begin{equation}
\epsilon_{sky} = 0.24 + 2.98 \times 10^{-8} e^{2}_{0}~exp\left(\frac{3000}{T_{amb}}\right)
\end{equation}

where $e_0$ is given by \cite{bib:bolton}:

\begin{equation}
e_0 = 6.11 \times 10^{\left(\frac{7.5T_d}{237.7 + T_d}\right)}
\end{equation}

and the sky radiance, $R_{sky}$, of a clear sky in Wm$^{-2}$ is:

\begin{equation}
R_{sky} = \epsilon_{sky}\sigma T_{amb}^4
\end{equation}

where $\sigma$ is the Stefan-Boltzmann constant. Similarly, incoming radiation from surrounding objects obscuring the night sky will have an effect, but this is disregarded here as it is assumed all mirrors will observe a completely unobscured night sky. This is a limitation of the present study, since this will not be true for all mirrors on a telescope, but nonetheless the majority should be unobscured. In this case:

\begin{equation}
R_{net} = R_{obj}-R_{sky} 
\end{equation}

where:

\begin{equation}
R_{obj} = \epsilon_{obj}\sigma T_{obj}^4 \label{eq:Robj}
\end{equation}

and $\epsilon_{obj}$ is the emissivity of the mirror in the 8 to 14 $\mu$m waveband.

\subsection{Net Effect of Thermal Processes}

Combining conduction, convection and radiative heat transfer, the net loss of thermal energy, $P_{net}$, from the front surface of a mirror of thickness $L$ exposed to a clear night sky is given by:

\begin{equation}
P_{net} = R_{net}-h(T_{amb}-T_{obj})+\frac{k}{L}(T_{back}-T_{front})
\end{equation}

where $T_{front}$ and $T_{back}$ are the temperatures of the back and front of the mirrors in degrees Kelvin.

\section{Outside Testing of Mirrors}

The mirrors used in the tests described here are listed in Table~\ref{table_mirrors}. These mirrors were tested by placing them outside at two locations in the UK, Durham (in the North of England) and Chelmsford (in the South) on nights which were as clear and calm as possible during December 2011 - March 2012. The mirrors were placed facing directly upwards and their surface temperatures were measured using K-type thermocouples attached to the mirrors' surfaces. These temperatures were recorded using simple remote data-loggers and the mirrors were also checked visually for the formation of condensation and/or frost during the tests.

\begin{table*}[h]
\begin{center}
\begin{tabular}{|c|c|c|c|c|c|}
\hline Mirror & Type & Materials & Surface & Size (m) & Thickness (m) \\ \hline
A36   & Solid  & Soda lime float glass & Al/SiO$_2$ & 0.6 & 0.02 \\
T191 & Solid & Kavalier SIMAX glass & Al/SiO$_2$ & 0.6 & 0.015 \\
T6086 & Solid & Kavalier SIMAX glass & Dielectric, SiO$_2$ and Ta$_2$O$_5$ & 0.6 & 0.015 \\
T1010 & Solid & Kavalier SIMAX glass & Al/SiO$_2$ + hydrophobic coating & 0.6 & 0.015 \\
K1 & Composite & Glass + open metal substrate & Al/SiO$_2$ & 0.4 & 0.08 \\
D1 & Composite & Alanod + Al honeycomb & Al + proprietary coating & 0.48 & 0.03 \\
M1 & Composite & Al surface + honeycomb & Al/SiO$_2$ & 0.17 & 0.025 \\
M2 & Composite & Glass + Al honeycomb & Al/SiO$_2$ & 0.2 & 0.024 \\
\hline
\end{tabular}
\caption{Sample Cherenkov telescope mirrors used for outdoor tests. Note that the K1 and D1 mirrors are hexagonal in shape, with the size measured from flat side to flat side, and the M1 and M2 mirrors are square. All the other mirrors are circular.}
\label{table_mirrors}
\end{center}
\end{table*}

\begin{table*}[h]
\begin{center}
\begin{tabular}{|c|c|c|c|c|c|c|c|c|}
\hline Date & A36 & T191 & T6086 & T1010 & K1 & D1 & M1 & M2 \\ \hline
18/12/11 & F & F & F & F & - & - & - & - \\
22/12/11 &   &   & C &   & - & - & - & - \\
28/12/11 &   &   &   &   & - & - & - & - \\
02/01/12 &   &   &   &   & - & - & - & - \\
05/01/12 &   &   & F &   & - & - & - & - \\
10/01/12 &   &   & C &   & - & - & - & - \\
12/01/12 &   &   & C &   & - & - & - & - \\
13/01/12 &   &   & F &   & - & - & - & - \\
16/01/12 &   &   & F &   & - & - & - & - \\
23/01/12 &   &   & C &   & - & - & - & - \\
02/02/12 &   & - & F &   &   &   & - & - \\
05/02/12 &   & - & F &   & C & C & - & - \\
07/02/12 &   & - & C &   &   &   & - & - \\
08/02/12 &   & - & F &   &   &   &   & - \\
15/02/12 &   & - & C &   &   &   &   & - \\
16/02/12 &   & - & C &   &   &   &   & - \\
24/02/12 &   & - &   &   &   &   &   & - \\
25/02/12 &   & - & C &   &   & C &   & - \\
28/02/12 &   & - & C & - &   &   &   & - \\
29/02/12 &   & - & C & - &   &   &   & - \\
01/03/12 &   & - & C & - &   &   &   &   \\
02/03/12 & - & - & C & - &   &   &   &   \\
05/03/12 & - & - & F & - &   &   &   &   \\
07/03/12 & - & - & - & - &   &   &   &   \\
10/03/12 & - & - & C & - &   &   &   &   \\
\hline
\end{tabular}
\caption{The results of testing sample mirrors outside. F denotes frost formation was observed on the mirror surface, C denotes condensation was observed, a blank column denotes that neither condensation nor frost was observed and a - denotes that the mirror was not tested on that occasion. Date of observation corresponds to the start date.}
\label{table_results}
\end{center}
\end{table*}

\begin{figure*}[!t]
  \centering
  \includegraphics[width=\textwidth]{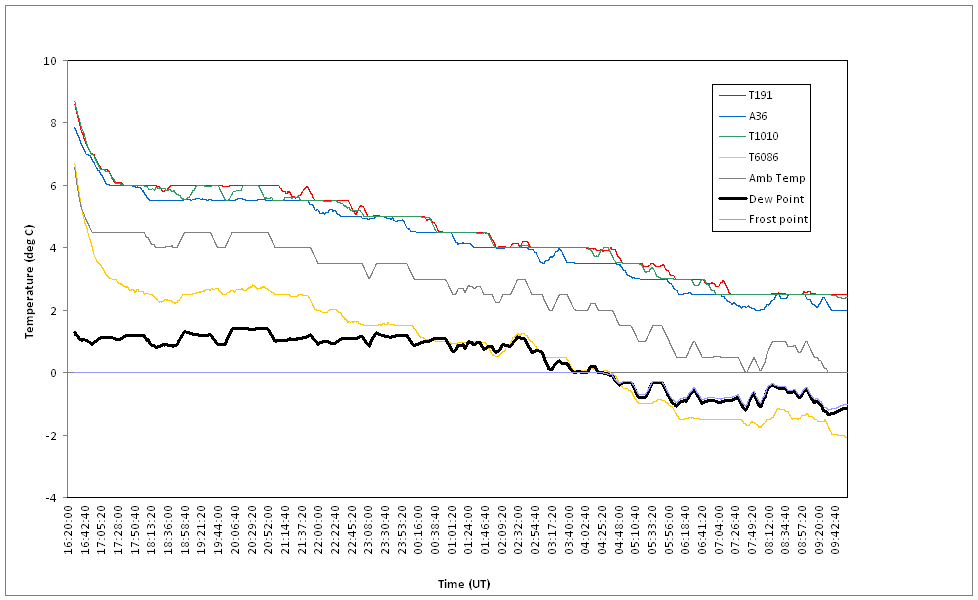}
  \caption{Temperatures of mirrors recorded on January 12th 2012; uncertainties are estimated at $\pm 0.5~^{\circ}C$. The temperature of T6086 (yellow line), the dielectric-coated mirror, falls much more rapidly than the other mirrors, which all have aluminium/quartz coatings. By $\sim 00:00$~UT the mirror temperature has fallen below the dew point (thick black line), allowing condensation to form, and by $\sim 04:30$~UT it has fallen below the frost point (dark blue line). The other mirrors (light blue, red and green lines) all remain a degree or two above the ambient temperature (grey line).}
  \label{temp_fig}
 \end{figure*}

Table~\ref{table_results} shows the results of testing the mirrors for condensation and the formation of frost on their surfaces (note that not all mirrors were available throughout the full study). These results show that the composite mirrors do show a slightly greater tendency to suffer condensation or frost than the solid glass mirrors, but this brief study suggests it is not a major effect. A much greater tendency to suffer condensation is observed in the case of the solid glass mirrors with the dielectric coating. The reason for these results can be seen in Figure~\ref{temp_fig}; the dielectric-coated mirror cools much more rapidly than the other glass mirrors and is therefore more likely to reach the dew point than the other mirrors on any given night.

\section{Discussion}

The outside mirror testing showed the initially surprising result that the solid glass dielectric-coated mirrors showed the greatest tendency of any of the mirror types to suffer condensation or frost, including the composite mirror types. It might be expected that the composite mirror types would suffer more from condensation due to their relatively low thermal conductivity $k$, which has an effect on the net loss of thermal energy via conduction, as shown in Equation~\ref{eq:cond}. However, another factor is clearly having a much greater effect, and this is the emissivity of the mirrors, $\epsilon_{obj}$, which is important for the radiative cooling of the mirrors, as shown in Equation~\ref{eq:Robj}.

Measurements of the emissivity of three of the mirrors, A36, T6086 and D1, showed that the emissivity of A36 and D1, both of which have aluminium front surfaces, is $\sim 0.09$ between 8 and 14 $\mu$m. However, the emissivity of T6086, the dielectric mirror, is $\sim 0.9$ over the same waveband. This will result in rapid cooling of the mirrors and hence the formation of condensation or frost, as observed in our trials.

\section{Conclusions}

We have shown that emissivity between 8 and 14 $\mu$m is the most important factor controlling the formation of condensation on Cherenkov telescope mirrors, not the method of construction. This is a particular issue for mirrors with dielectric coatings which have been tuned to provide low reflectivity in the red region of the spectrum to reduce the effects of the night sky background. The corollary of this cut-off beyond $\sim 600$ nm is high emissivity in the mid-infrared, which causes the mirrors to cool quickly. It may be possible to improve these coatings in order to reduce this effect.

The equations which describe surface cooling can be used to predict the surface temperatures of mirrors, provided that the basic mirror and meteorological data are available, and comparison of predicted mirror temperature with the dew point makes it possible to estimate when mirrors of a given type will suffer condensation at a particular candidate telescope site. This is an important consideration for the design of Cherenkov telescope mirrors and the choice of telescope site and will be the subject of further study.

\vspace*{0.5cm}
\footnotesize{{\bf Acknowledgment:}{We would like to acknowledge colleagues in CTA for the provision of mirror samples, and Prof. David Wood of the School of Engineering and Computer Sciences at Durham University for help with emissivity measurements. We gratefully acknowledge support from the agencies and organizations listed in this page: http://www.cta-observatory.org/?q=node/22}}

\end{document}